\documentclass[pdflatex,sn-nature]{sn-jnl}

\usepackage{graphicx}%
\usepackage{amsmath}%
\usepackage{booktabs}%
\usepackage[version=4]{mhchem}
\usepackage{siunitx}
\usepackage{float}
\usepackage{anyfontsize}

\begin{document}

\title{An interstellar energetic and non-aqueous pathway to peptide formation}

\author[1*]{Alfred Thomas Hopkinson}
\author[1]{Ann Mary Wilson}
\author[1]{Joe Pitfield}
\author[2]{Alejandra Traspas Muiña}
\author[3]{Richárd Rácz}
\author[3]{Duncan V. Mifsud}
\author[3]{Péter Herczku}
\author[3,4,5]{Gergő Lakatos}
\author[3]{Béla Sulik}
\author[3]{Zoltán Juhász}
\author[3]{Sándor Biri}
\author[6]{Robert W. McCullough}
\author[6,7]{Nigel J. Mason}
\author[8]{Carsten Scavenius}
\author[1,9]{Liv Hornekær}
\author[1**]{Sergio Ioppolo}

\affil[1]{Center for Interstellar Catalysis, Aarhus University, Physics and Astronomy Department, Aarhus C, 8000, Denmark}

\affil[2]{School of Electronic Engineering and Computer Science, Queen Mary University of London, London E1 4NS, United Kingdom}
\affil[3]{HUN-REN Institute for Nuclear Research (Atomki), Debrecen H-4026, Hungary}
\affil[4]{Institute of Chemistry, University of Debrecen, Debrecen H-4032, Hungary}
\affil[5]{Doctoral School of Chemistry, University of Debrecen, Debrecen H-4032, Hungary}
\affil[6]{Department of Physics and Astronomy, School of Mathematics and Physics, Queen’s University Belfast, Belfast BT7 1NN, United Kingdom}
\affil[7]{Centre for Astrophysics and Planetary Science (CAPS), School of Physics and Astronomy, University of Kent, Canterbury CT2 7NH, United Kingdom}
\affil[8]{Department of Molecular Biology and Genetics - Protein Science, Aarhus University, Aarhus, 8000, Denmark}
\affil[9]{Interdisciplinary Nanoscience Center, Aarhus University, Aarhus C, 8000, Denmark}

\affil[*]{a.hopkinson@phys.au.dk}
\affil[**]{s.ioppolo@phys.au.dk}

\abstract{The origin of the molecular building blocks of life is a central question in science. A few $\alpha$-amino acids such as glycine, the simplest proteinogenic amino acid, have been detected in meteorites and comets, indicating an extraterrestrial origin for some prebiotic molecules. However, the formation of peptides, short chains of $\alpha$-amino acids linked by peptide bonds, under astrophysical conditions has remained unresolved. Here we show that the building blocks of proteins can form in interstellar ice analogues exposed to ionising radiation, without the presence of liquid water. Using isotopically labelled glycine irradiated with protons at cryogenic temperatures, we detect the formation of glycylglycine, the simplest dipeptide, along with deuterated and nondeuterated water as by-products. Peptide bond formation is confirmed by infrared spectroscopy and high-resolution mass spectrometry, which also reveal the production of other complex organic species. These findings demonstrate a non-aqueous route to peptide formation under space-like conditions and suggest that such molecules could form in the cold interstellar medium and be incorporated into forming planetary systems. Our results challenge aqueous-centric models of early biochemical evolution and broaden potential settings for the origins of life.}


\maketitle

\section{Main text}

\begin{figure}[H]
    \centering
     \includegraphics[width=100mm]{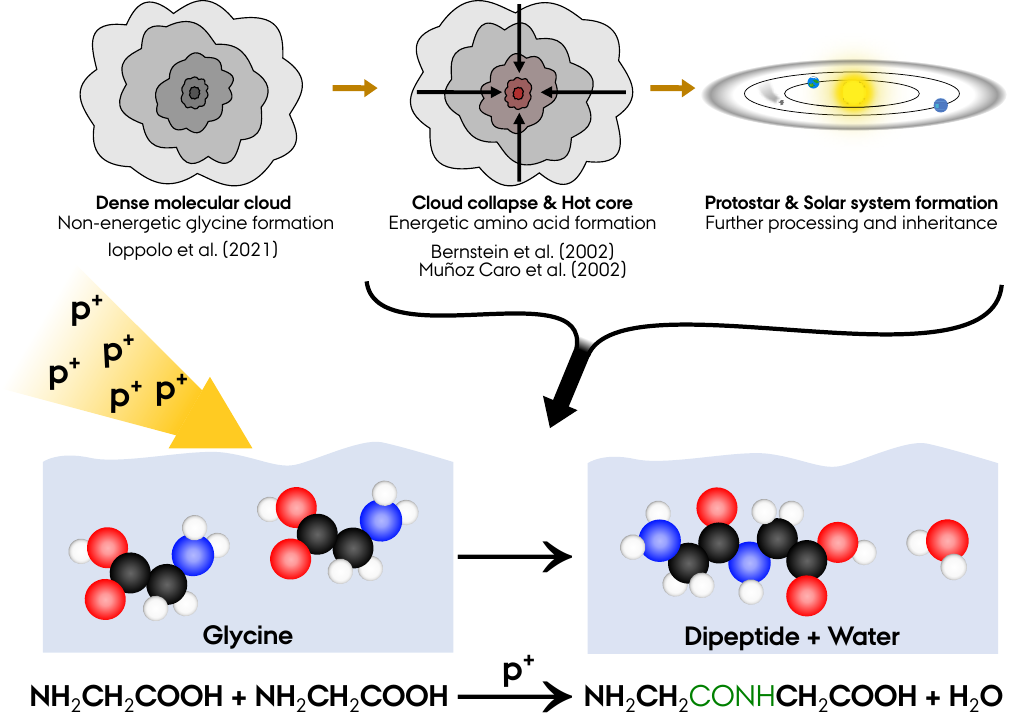}
     \caption{Peptide formation induced by energetic processing in the interstellar medium. \textbf{Top panel:} Schematic representation of the current understanding of stellar evolution, with emphasis on the stages relevant to amino acid formation. Existing studies that have identified potential synthetic pathways for amino acids during these evolutionary phases are indicated for reference \cite{2021NatAs...5..197I,Uvamino,anotherUVamino}. The present study focuses predominantly on the final two stages of stellar evolution, where increased energetic processing is prevalent. \textbf{Bottom panel:} Proposed reaction pathway illustrating the ion irradiation of glycine to form glycylglycine, the simplest dipeptide, accompanied by water formation, as investigated in this study. The process is presented both schematically as interstellar ice grains containing glycine molecules and as a balanced chemical equation, pointing out the role of energetic processing in peptide-bond formation (highlighted in green) under astrophysical conditions.}
     \label{Graph}
 \end{figure}

The emergence of life on Earth has been linked to the availability of prebiotic molecules in the early solar system, yet the origin and evolution of these essential compounds remain open questions \cite{2022FrASS...8..255P}. Some $\alpha$-amino acids, the fundamental building blocks of proteins, such as glycine (\ce{NH2CH2COOH}), have been detected in meteorites and comets, suggesting that biologically relevant molecules form and evolve in space before being delivered to planetary surfaces \cite{doi:10.1021/cr2004844,2005RSPTA.363.2729S,2019A&A...630A..32H, 2015Sci...347a0628C,wild2,morewild}. Laboratory experiments have also shown that glycine can form through nonenergetic processes in astrophysical ice analogues, providing plausible formation pathways under dense molecular cloud conditions \cite{2021NatAs...5..197I} (Fig.~\ref{Graph}). However, the subsequent chemical evolution of glycine under astrophysical energetic processing, such as UV photolysis, shock processing, and cosmic ray irradiation, remains poorly understood, particularly with regard to the formation of peptide-bonded species \cite{mate,2014FaDi..168..267M,protom,2014MNRAS.441.3209P,10.1093/mnras/staa3939,shock,singh2020shock}. Biological peptides, chains of $\alpha$ amino acids, play key roles in catalysis, structural organisation, and protocellular membrane formation. 

Previous studies have shown that energetic processing of interstellar ice analogues produces complex organic molecules (COMs), including amino acids \cite{Uvamino, anotherUVamino}, and that subsequent irradiation can drive both amino acid degradation and amide bond formation \cite{mate,2014FaDi..168..267M,protom,2014MNRAS.441.3209P,10.1093/mnras/staa3939}. However, the clear detection of glycylglycine (\ce{NH2CH2CONHCH2COOH}), the simplest biological dipeptide, under such conditions remains elusive. A recent study proposed a nonenergetic route to non-biological peptide formation via polymerisation of aminoketene (\ce{NH2CHCO}) under astrophysically relevant conditions \cite{2022NatAs...6..381K}. Although constrained by the limited atomic carbon in dense molecular clouds, this mechanism bypasses amino acid formation and may represent an additional pathway to peptide-bond synthesis during the early stages of star formation, from translucent to dense cloud phases \cite{1995Sci...270.1455S}. 

Here, we investigate the energetic processing of glycine under astrophysically relevant conditions representative of advanced stages of star formation, including cloud collapse, protoplanetary disks, and nascent planetary environments. Isotopically labelled glycine ices at 20~K were irradiated with protons spanning 10--1000~keV to simulate the effects of galactic cosmic rays and stellar winds, probing the formation of higher-order COMs. Our results provide evidence for an interstellar solid-phase peptide formation mechanism driven by ionising radiation (Fig.~\ref{Graph}). These findings indicate that glycine formed on icy grains can evolve into more complex biomolecular structures before planetary accretion. This demonstrates that peptide synthesis, a key step toward biological polymers, can occur under non-aqueous conditions \cite{doi.org/10.1002/cplu.202300492}. Such results broaden the potential environments for abiogenesis, from the cold, dense ISM to solar system conditions, and support the view that life’s chemical precursors may have extraterrestrial origins.

\section{Molecule formation through irradiation}
\begin{figure}[H]
    \centering
     \includegraphics[width=100mm]{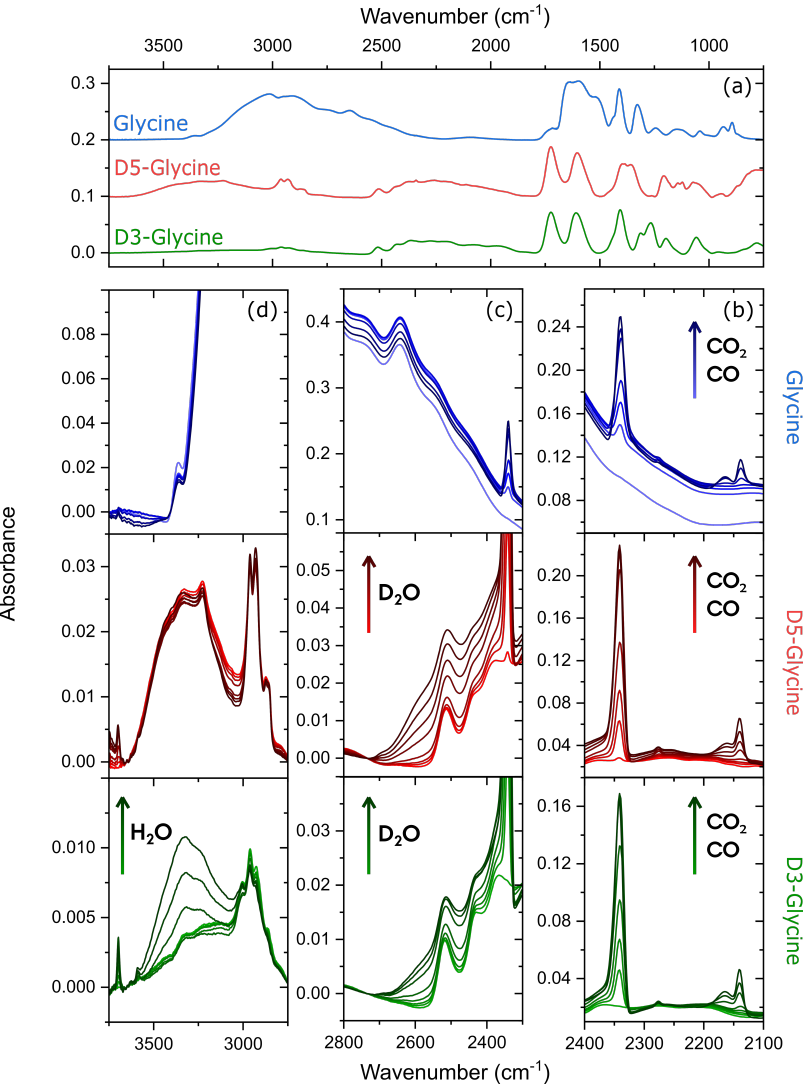}
     \caption{FTIR spectra of the three glycine isotopologues and their changes during 10~keV proton exposure. \textbf{a)} FTIR spectra of glycine (blue), fully deuterated D5-glycine (red), and partially deuterated D3-glycine (green) deposited onto 20~K ZnSe windows. Spectra are offset and scaled for clarity. \textbf{b)}, \textbf{c)}, and \textbf{d)} FTIR spectra from 2400--2100, 2800--2300, and 3750--2750~cm$^{-1}$ ranges, respectively, for all three different glycine isotopologues acquired as a function of 10~keV proton irradiation time. The darker the colour, the more irradiated the sample to an end point of a proton fluence of \qty{6E15}{H^+cm^{-2}}. \textbf{b)} \ce{CO2}, \ce{OCN-}, and \ce{CO} formation (peaks centred at 2341, 2165, and 2140~cm$^{-1}$, respectively) from the irradiation of all three glycine isotopologues. \textbf{c)} \ce{D2O} formation for the D5-glycine and D3-glycine experiments. \textbf{d)} \ce{H2O} formation largely for the D3-glycine irradiation.}
     \label{Dep+irr}
 \end{figure}

We aim to demonstrate that irradiation of glycine isotopologues by protons leads to the formation of peptides and the by-product of water (\ce{H2O}) and fully deuterated water (\ce{D2O}), their production being dependent on isotopic substitution. Such findings provide valuable insight into the chemical evolution of glycine and water-bearing species in astrophysical environments. Fig.~\ref{Dep+irr} presents the infrared (IR) spectra of three isotopologues of glycine subjected to energetic processing by 10~keV proton irradiation at 20~K. The darker the spectra, the greater the extent of proton bombardment up to a proton fluence of \qty{6E15}{H^+cm^{-2}}. The studied isotopologues include nondeuterated glycine (\ce{NH2CH2COOH}), partially deuterated D3-glycine (\ce{ND2CH2COOD}), where deuterium replaces hydrogen in the amine and carboxyl groups, and fully deuterated D5-glycine (\ce{ND2CD2COOD}). Fig.~\ref{Dep+irr}a displays the deposition spectra of these isotopologues at 20~K. The presence of neutral glycine is indicated by strong absorption bands at 1730 and 1240~cm$^{-1}$. Previous studies have shown that the Fourier Transform Infrared (FTIR) spectrum of glycine varies with factors such as substrate temperature, molecular deposition energy, and layer thickness \cite{mate,howdep}. In the present study, deposited glycine exists as a mixture of neutral and zwitterionic forms, the latter characterised by a positively charged amine (\ce{NH3+}) and a negatively charged carboxylate (\ce{COO-}) group. In particular, significant spectral differences arise between deuterated and nondeuterated glycine, with the most prominent being the absence of broad absorption around 3000~cm$^{-1}$ associated with the \ce{NH} and \ce{CH} stretching modes and the emergence of enhanced absorption near 2250~cm$^{-1}$ attributed to the \ce{ND} and \ce{CD} stretching modes \cite{howdep,Kaiser_2013}. 

Fig.~\ref{Dep+irr}b highlights the absorption features associated with the products formed during irradiation, within the 2400--2100~cm$^{-1}$ range. A sharp peak at 2341~cm$^{-1}$ corresponds to carbon dioxide (\ce{CO2}), produced by glycine decarboxylation \cite{Gerakines2012InSM}. The peak at 2140~cm$^{-1}$, assigned to carbon monoxide (\ce{CO}), arises from the irradiation-induced decomposition of \ce{CO2}, which becomes prominent at higher irradiation doses, consistent with previous findings \cite{Gerakines2012InSM,JHEETA2012208}. Furthermore, a broad feature centred at 2165~cm$^{-1}$ has been attributed to the cyanate ion (\ce{OCN-}), although contributions from larger COMs containing similar functional groups are likely, as discussed in the Supplementary Information and in Ref.~\cite{mate}. The intensities of all of these absorption bands increase with prolonged irradiation. Fig.~\ref{Dep+irr}c and \ref{Dep+irr}d focus on spectral regions indicative of \ce{H2O} and \ce{D2O} formation. Fig.~\ref{Dep+irr}c further examines the 2800--2300~cm$^{-1}$ range, where the formation of \ce{D2O} is identified through the stretching modes \ce{O-D} \cite{OD}. Nondeuterated glycine exhibits minimal spectral changes across this range, while both D3-glycine and D5-glycine show a significant increase in absorption with irradiation, confirming the formation of \ce{D2O}. Interestingly, while D5-glycine maintains a continuous increase in absorption, D3-glycine exhibits saturation at later stages of irradiation. The \ce{H2O} absorption band, centred around 3350~cm$^{-1}$ and shown in Fig.~\ref{Dep+irr}d, corresponds to the \ce{O-H} stretching modes \cite{PALUMBO201064,OD}. For nondeuterated glycine, detection of \ce{H2O} formation is challenging due to strong intrinsic absorptions in the same spectral range. In Supplementary Fig.~1a where 20~K glycine is processed more thoroughly by 1 MeV protons, water formation is observable. However, partially deuterated D3-glycine exhibits an increasing absorption feature in this region, indicating the formation of \ce{H2O} or partially deuterated water (\ce{HDO}), particularly in the final stages of irradiation. Fully deuterated D5-glycine shows no significant changes in this region. 

\begin{figure}[H]
    \centering
     \includegraphics[width=100mm]{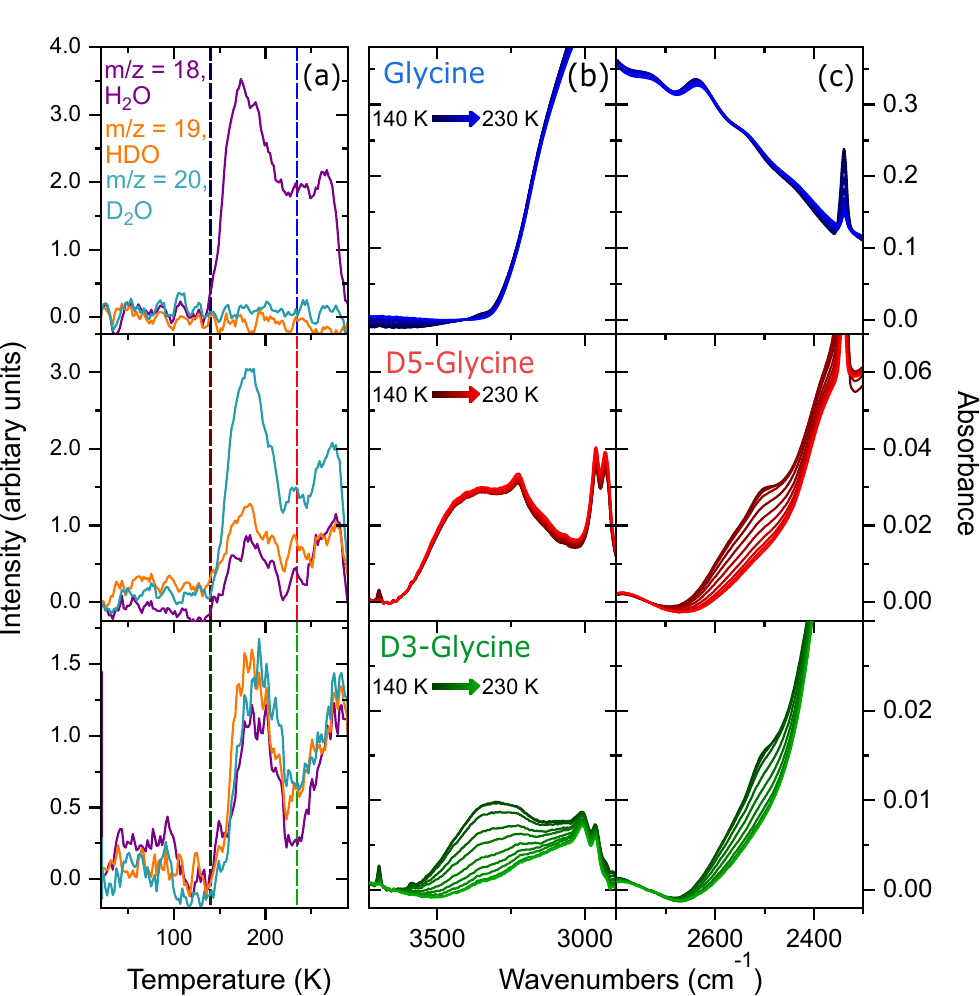}
     \caption{QMS detections of \ce{H2O},\ce{HDO}, and \ce{D2O} with the corresponding FTIR spectra during TPD measurements of all three glycine isotopologues after proton exposure. \textbf{a)} Temperature programmed desorption spectra of irradiated glycine, D5-glycine, and D3-glycine from 20--280~K. The traces are for m/z = 18 - \ce{H2O}, m/z = 19 - \ce{HDO}, and m/z = 20 - \ce{D2O}, showing broad desorption from 140--230~K as highlighted by the vertical dashed lines. \textbf{b)} and \textbf{c)} The evolution of the FTIR spectra with increasing temperature during the TPD from 3750--2750 and 2800--2300~cm$^{-1}$, respectively, for all three different glycine isotopologues. The brighter the colour, the higher the sample temperature.}
     \label{TPD}
 \end{figure}

The formation of \ce{H2O} and \ce{D2O} is further corroborated by complementary temperature-programmed desorption (TPD) measurements performed at the end of the irradiation by heating the ice at a constant rate up to room temperature (Fig.~\ref{TPD}, see Methods). Fig.~\ref{TPD}a presents quadrupole mass spectrometry (QMS) traces for m/z = 18 (\ce{H2O}), m/z = 19 (\ce{HDO}) and m/z = 20 (\ce{D2O}), showing broad desorption between 170--220~K. The IR spectra recorded during TPD, shown in Figs.~\ref{TPD}b and \ref{TPD}c, confirm these findings by highlighting any decrease in the intensity of the water spectral bands as a function of increasing temperature. The lighter the spectra, the higher the ice temperature. For nondeuterated glycine, substantial \ce{H2O} desorption is evident in the QMS data without traces of desorption of \ce{HDO} and \ce{D2O}. However, the corresponding IR spectral changes remain subtle because of overlapping glycine absorption bands. For D5-glycine, clear desorption of \ce{D2O} (m/z = 20) is observed in the QMS and IR data, with a reduction in the intensity of the \ce{O-D} band in Fig.~\ref{TPD}c, while minor formation of \ce{HDO} and \ce{H2O} is detected only in the QMS. In contrast, D3-glycine shows desorption of \ce{H2O}, \ce{HDO}, and \ce{D2O} in both QMS and IR data, with corresponding reductions in the \ce{O-H} and \ce{O-D} absorption bands at 3300 and 
2550~cm$^{-1}$ (Figs.~\ref{TPD}b and \ref{TPD}c).

\section{Water formation through peptide-bond formation}\label{Water}

During glycine proton bombardment, \ce{H2O} can form through a two-step reaction pathway that involves glycine irradiation products. Although exact mechanisms are unclear, one hypothesis suggests that a radical oxygen atom, produced during the proton-induced conversion of \ce{CO2} to \ce{CO} (Ref.~\cite{JHEETA2012208}), reacts with hydrogen-containing molecules to generate \ce{H2O}. Alternatively, water could form through a single-step condensation reaction leading to the formation of a peptide bond (\ce{R_1-CONH-R_2}). This process involves the reaction of the carboxyl group of one glycine molecule with the amine group of another to form an amide bond, releasing water in the process, as shown in Eq.~(\ref{pepform}), where \ce{R1} and \ce{R2} represent the rest of the molecule. 

\begin{equation}\label{pepform}
    \ce{R_1-COOH} + \ce{NH2-R_2} \rightarrow \ce{R_1-CONH-R_2} + \ce{H2O}
\end{equation}

The formation of deuterated water \ce{D2O} upon bombardment of deuterated glycine follows the same pathways as described above for \ce{H2O}. However, the presence of deuterium leads to the reaction shown in Eq.~(\ref{pepformD5}) for both fully deuterated (D5) glycine and partially deuterated (D3) glycine.

\begin{equation}\label{pepformD5}
     \ce{R_1-COOD} + \ce{ND2-R_2} \rightarrow \ce{R_1-COND-R_2} + \ce{D2O}
\end{equation}

\begin{figure}[H]
    \centering
     \includegraphics[width=100mm]{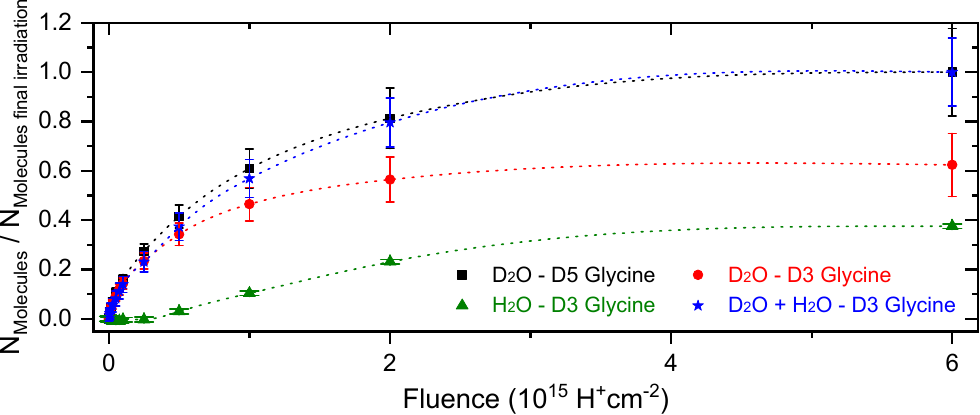}
     \caption{Number of \ce{D2O} and \ce{H2O} molecules normalised to the final irradiation yield plotted as a function of the fluence of proton bombardment, showing the rate of formation. The number of \ce{D2O} molecules formed during irradiation of D5-glycine (black squares) . \ce{H2O} is not formed during D5-glycine processing. In the case of D3-glycine, the sum of the amount of \ce{D2O} and \ce{H2O} is shown (blue stars) and the separate components of \ce{H2O} (green triangles) and \ce{D2O} (red circles) are normalised to this total. The \ce{H2O} error bands are the maximum spread of the change of the baseline and the \ce{D2O} error bands are the standard deviation of the different fits of the \ce{D2O} reference spectrum. Lines are added to guide the eye.}
     \label{waterlevel}
 \end{figure}

The evolution of the differences in the isotopic composition of the water produced indicates that different processes contribute to the formation of water. This is illustrated in Fig.~\ref{waterlevel}, where the number of \ce{D2O} and \ce{H2O} molecules, normalised to the total number of water molecules at the end of the irradiation, is plotted against the fluence of proton bombardment. 

The amount of \ce{H2O} is determined by measuring the area $A$ of the water absorption band in the 3610--3065~cm$^{-1}$ region. The number of molecules is then calculated using equation $N_{molecules} = \frac{A\times \text{ln}(10)}{B}$, where $B$ represents the band strength of water, taken as $2.85\times10^{-16}$~cm~molecules$^{-1}$ (mean value of the band strengths at 10 and 30 K, Ref~\cite{newwater}). The amount of \ce{D2O} is determined by scaling a reference band from Ref~\cite{2018MNRAS.479..130U} to several points of the spectrum, averaging the area of the band scaled to each point, from which the number of molecules is calculated. This procedure allows for an estimate of the amount of \ce{D2O} formed in D3-glycine and D5-glycine, and \ce{H2O} formed in D3-glycine as a function of the irradiation dose. The \ce{H2O} error bands are the maximum spread of the change of the baseline in this region and the \ce{D2O} error bands are the standard deviation of the different fits of the \ce{D2O} reference spectrum. The irradiation dose was chosen to approximate typical solar wind conditions, with proton fluxes up to 10 keV estimated at approximately \qty{E10}{H^+cm^{-2}s^{-1}}, depending on solar activity and distance from the Sun \cite{SZNAJDER20234923}. 

The data reveal that the formation rates of \ce{D2O} differ between D3- and D5-glycine. In D3-glycine, \ce{D2O} production is relatively faster at lower fluence than in D5-glycine. Furthermore, \ce{H2O} formation is initially slow in D3-glycine, increasing as irradiation progresses. These trends are also evident in Fig.~\ref{Dep+irr}, where the formation of \ce{H2O} occurs only at higher irradiation fluences for D3-glycine, while the production of \ce{D2O} increases more gradually from the beginning. This shows that there are two distinct reaction pathways for the formation of \ce{D2O} and \ce{H2O}, with \ce{H2O} primarily formed through a two-step reaction at higher proton fluences when there are more glycine degradation products present, namely \ce{CO} and \ce{CO2}. In D3-glycine, which contains hydrogen in the alpha carbon position, \ce{H2O} is likely to be formed via the two-step reaction pathway discussed above, rather than through the formation of peptide bonds. In contrast, \ce{D2O} in D3-glycine is mainly produced by peptide formation, as described in Eq.~(\ref{pepformD5}). For D5-glycine, \ce{D2O} is formed through a combination of peptide formation and two-step processes, which explains the continued production of \ce{D2O} even at high proton fluences. When the amounts of \ce{D2O} and \ce{H2O} for D3-glycine are combined and then normalised to the total amount at the end of the irradiation, their formation rate profile becomes more similar to that of \ce{D2O} in D5-glycine, showing that glycine proton bombardment leads to both peptide-bond formation and two-step reaction processes that contribute to water production. To investigate the effects of galactic cosmic rays on interstellar ice grains, glycine was irradiated with 1 MeV protons. The resulting evolution of water formation under this higher-energy processing was similar to the 10 keV proton processing with the corresponding results presented in Supplementary Fig.~1b.

\section{Amide bands assignment}
\begin{figure}[H]
    \centering
     \includegraphics[width=100mm]{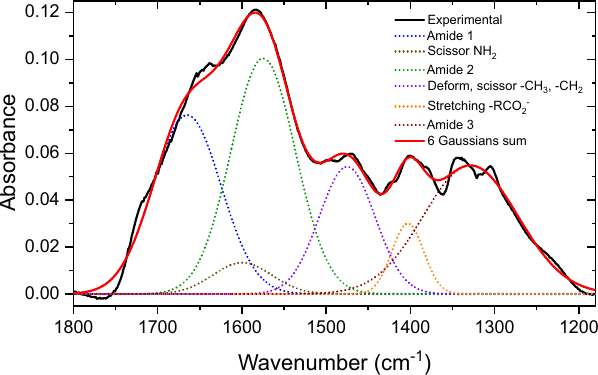}
     \caption{FTIR spectra of the residue obtained after processing glycine with 1~MeV protons with a fluence of \qty{1.24E15}{H^+cm^{-2}}. The spectrum of the residue is fitted with six Gaussians of different IR bands as found in literature \cite{mate,2022NatAs...6..381K}. The Amide 1, 2, and 3 bands are indicators of peptide-bond formation after energetic processing of glycine \cite{Kaiser_2013}.}
     \label{amidebonds}
\end{figure}

Irradiation with 1~MeV protons simulating the effect of galactic cosmic rays leads to more extensive molecular processing compared to 10~keV experiments, which resemble the stellar wind more closely. Consequently, at the end of the irradiation, fewer IR spectral features of intact glycine remained. This processing resulted in the formation of \ce{H2O}, with its IR spectrum presented in the Supplementary Fig.~1a. As in the 10~keV experiments, the formation of \ce{H2O} is attributed to both the formation of the peptide bond and the two-step reaction mechanisms. Furthermore, irradiation with 1~MeV protons left a visible residual film on the window, which is subsequently analysed \textit{in situ}. The corresponding IR spectrum, deconvolved into six Gaussian components, is shown in Fig.~\ref{amidebonds}, while the Supplementary Fig.~2 presents an alternative fit using five Gaussians. These Gaussian positions were chosen to reflect the bands expected to be present for COMs that contain amide bands such as peptides, as shown in other work \cite{mate,2022NatAs...6..381K,Kaiser_2013}. Complementing the assignment of the band in the literature were also the calculated spectra of glycylglycine shown in the Supplementary Fig.~3 that supported the assignment of the band here due to its possible formation. The residue is fit with five and six bands to compare the differences between previous assignments with the six Gaussian fitting slightly better and containing the scissor \ce{NH2} mode likely to be present here. The central positions and assignments of these spectral bands are listed in Supplementary Table~1.

The selected bands are characteristic of the formation of the peptide bond and are consistent with residues observed in previous studies \cite{mate,2022NatAs...6..381K,Kaiser_2013}. In particular, the presence of Amide 1, 2, and 3 bands is indicative of peptide bonding within the processed sample. These features are expected, given the concurrent formation of \ce{H2O} during irradiation, which supports their assignment. Specifically, the Amide 1 band at 1665~cm$^{-1}$ arises from carbonyl (\ce{CO}) stretching, while the band at 1599~cm$^{-1}$ corresponds to the \ce{NH2} scissor mode \cite{2022NatAs...6..381K}. The Amide 2 band at 1573~cm$^{-1}$ is attributed to an out-of-phase combination of \ce{CN} stretching and \ce{NH} bending, while the Amide 3 band at 1325~cm$^{-1}$ corresponds to an in-phase combination of these vibrations \cite{mate,Kaiser_2013,2022NatAs...6..381K}. Additional spectral characteristics include the deformation mode of the deprotonated carboxylate groups (\ce{-RCO2$^-$}) around 1404~cm$^{-1}$ and the deformation and scissor modes of the methyl (\ce{-CH3}) and methylene (\ce{-CH2-}) groups at 1470~cm$^{-1}$ (Ref.~\cite{mate,Kaiser_2013}). The relative intensities of these fitted bands, and their relative widths are similar to those reported in literature \cite{mate}. Although the presence of amide 1, 2, and 3 bands supports peptide bond formation, we recognise that IR bands in this region can also arise from overlapping carbonyl-containing species such as carboxylic acids or ketones present in the residue and therefore are insufficient to definitively confirm the formation of peptide bonds. However, the presence of amide bands, in conjunction with the formation of water, provides strong \textit{in situ} evidence for the synthesis of peptide-bonded molecules under energetic processing of amino acids at low temperatures characteristic of dense interstellar clouds.

\section{\textit{Ex situ} analysis}
\begin{figure}[H]
    \centering
     \includegraphics[width=120mm]{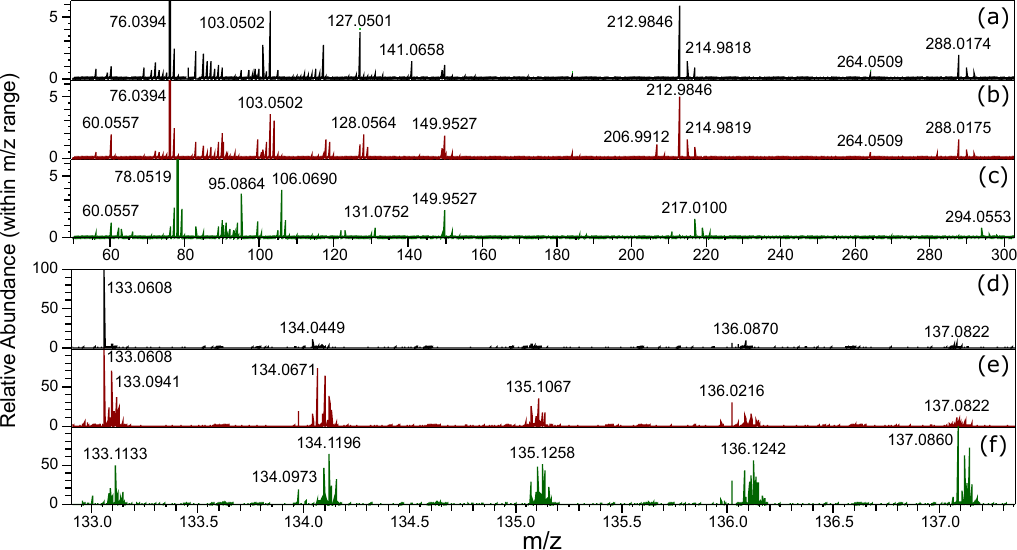}
     \caption{The \textit{ex situ} ESI-MS selected data on the newly detected molecules with some of the larger masses labelled. \textbf{a)} Relative abundances of total measured masses from the nondeuterated glycine, \textbf{b)} the D3-glycine, \textbf{c)} the D5-glycine. \textbf{d)}, \textbf{e)}, \textbf{f)} are the same as above but focused on an m/z range of 133 to 138 giving abundance relative to all molecules within that mass range. Non-processed glycine is detected by m/z = 76 for \textbf{a)} and \textbf{b)}, and by m/z = 78 for \textbf{c)}. Glycylglycine the glycine dipeptide is detected by m/z = 133 for \textbf{d)} and \textbf{e)}, and by m/z = 137 for \textbf{f)}.}
     \label{exsitu}
 \end{figure}

Fig.~\ref{exsitu} presents the mass spectra of glycine samples irradiated with 10~keV protons and analysed \textit{ex situ} using electrospray ionisation mass spectrometry (ESI-MS) after irradiation, as described in Methods. Fig.~\ref{exsitu}a corresponds to nondeuterated glycine, \ref{exsitu}b D3-glycine, and \ref{exsitu}c D5-glycine. The spectra indicate the continued presence of deposited glycine, with m/z = 76.04 for both nondeuterated glycine and D3-glycine. Although these samples initially contained different isotopologues, dissolution in solution facilitates proton exchange, resulting in identical detected masses. In contrast, for D5-glycine, the detected mass is m/z = 78.05, since all hydrogen atoms - except the two in the central alpha carbon - are expected to undergo exchange. Supplementary Fig.~4 provides full-scale spectra, illustrating the total glycine signal.

The analysis reveals a range of new molecular masses that extend to m/z = 300, substantially larger than the original glycine molecule. This indicates that energetic processing induces the formation of COMs. Of particular interest is investigating the potential formation of glycylglycine, the glycine dipeptide. This has a mass of 132.05 amu and its structure is shown in Supplementary Table~2. When in solution and protonated, the expected m/z value of this molecule would be 133.06 which is detected in both Fig.~\ref{exsitu}d and \ref{exsitu}e. As with glycine measured in these samples, dissolution in solution facilitated proton exchange, resulting in identical detected masses. However, fully deuterated D5-glycine will not exchange protons in central carbons and will then have an m/z of 137.09 which is also detected and shown in Fig.~\ref{exsitu}f. Together, these results give strong evidence that the energetic processing of glycine leads to the formation of glycylglycine.

The precise identities of the other products that contain amide and peptide bonds require further investigation; particular attention is paid to the species detected at m/z = 103.05 in the nondeuterated glycine sample. This mass is consistent with \textit{N}-formylglycinamide (\ce{NH2CH2CONHCHO}), whose resulting structure is shown in the Supplementary Table~2. \textit{N}-Formylglycinamide has a molecular weight of 102.04~amu, which, after protonation in solution, has an m/z = 103.05. \textit{N}-formylglycinamide is a subunit of formylglycinamide ribonucleotide amidotransferase (FGAR-AT) \cite{doi:10.1021/bi800329p}, an enzyme involved in purine biosynthesis, precursors of RNA and DNA, and in cellular energy metabolism \cite{PEDLEY2017141} so its formation in the ISM is of considerable interest. In deuterated samples, functional groups at the molecular termini undergo proton exchange, leaving only central deuterium atoms intact. In Fig.~\ref{exsitu}c, a peak appears at m/z = 106.07, indicating three retained deuterium atoms compared to the nondeuterated sample. This shows that deuterium is incorporated in the carbon and nitrogen of the peptide bond (\ce{NH2CD2CONDCHO}), which may stabilise against rapid exchange due to the adjacent aldehyde group. If this interpretation is correct, the expected m/z would be 106.07, which matches the experimental observations. To further investigate molecular structures, the fragmentation analysis of the ions is shown in Supplementary Fig.~5. The evidence for the presence of COMs, such as \textit{N}-Formylglycinamide, that contain \ce{OCN} and \ce{OCN-} groups is supported by an IR absorption feature at $\sim$2160~cm$^{-1}$ in Fig.~\ref{Dep+irr}b. This band persists above 200~K (Supplementary Figure~6), indicating that it is not only due to transient \ce{OCN-} species. A simulated IR spectrum of a related molecule exhibits a characteristic IR absorption in this region (Supplementary Fig.~7), which confirms the probability of \ce{OCN}-bearing COM formation during glycine irradiation. The direct detection of molecules that exceed the molecular weight of glycine is particularly significant because they may represent biologically relevant COMs.

\section{Discussion}
Although glycine has thus far been firmly detected only within the Solar System and not yet directly observed in the ISM, its presence is strongly supported by both laboratory simulations and astrochemical modelling. Experimental and theoretical studies have shown that glycine can be synthesised under astrophysically relevant conditions via nonenergetic surface chemistry, particularly through atom-addition reactions on icy grains \cite{2021NatAs...5..197I}, as well as through energetic processing mechanisms such as UV photolysis, cosmic ray bombardment, and shock-driven chemistry \cite{Uvamino,anotherUVamino,doi:10.1021/cr400153k}. Furthermore, aqueous alteration processes within meteorite parent bodies provide a complementary route for the formation of glycine during the early stages of planetary system evolution \cite{article}. While the precise abundance of glycine across astrophysical environments remains uncertain, the diversity of its formation pathways and the extended timescales over which such reactions can proceed suggest that glycine is a chemically plausible and potentially widespread molecule under space conditions \cite{2021NatAs...5..197I}.

In regions where thermal processing permits the sublimation of volatile ices, such as hot cores, protostellar environments, planetary surfaces, and meteorite-exposed bodies, amino acids such as glycine become directly subject to energetic irradiation on interstellar grains. This processing, whether by UV photons, electrons, or protons, predominantly leads to molecular fragmentation, producing simpler species such as \ce{CO}, \ce{CO2}, \ce{CN^-}, and \ce{OCN^-} \cite{mate,2014FaDi..168..267M,2014MNRAS.441.3209P,10.1093/mnras/staa3939}. However, irradiation also drives the assembly of more complex molecules. Previous studies have reported the emergence of infrared bands attributed to amide functional groups, which is indicative of the formation of peptide bonds \cite{mate,protom,2014MNRAS.441.3209P}. In the present work, we validate this process through the concurrent \textit{in situ} detection of both water and amide features, providing direct evidence that energetic processing of glycine facilitates peptide bond synthesis. Complementary \textit{ex situ} analyses further reveal the formation of larger COMs that carry amide and peptide linkages. Notably, the identification of the simplest dipeptide under such simulated conditions highlights the potential for classical proteinogenic structures to emerge abiotically in space at cold (20~K) temperatures. Additional complexity may arise under irradiation at higher temperatures, where enhanced molecular diffusion could enable further reaction pathways. The tentative detection of a second compound, \textit{N}-formylglycinamide, further supports the conclusion that chemically diverse amide-bearing species can form under astrophysical conditions.

Peptides are known to be efficient catalysts for amino acid derivatives and amino acid synthesis \cite{Gorlero2009153}. Thus, if they are present in space, once delivered to planetary surfaces, they could facilitate further polymerisation, contributing to prebiotic chemistry. This autocatalytic behaviour may have played a role in the early molecular evolution on Earth, allowing natural selection at the molecular level \cite{doi:10.1080/07391102.2022.2088619}. Peptides may have played a role in early structural and catalytic functions by contributing to the formation of protocells and other essential structures for life \cite{https://doi.org/10.1002/anie.201000212,doi:10.1098/rsif.2021.0641}, further highlighting their potential prebiotic relevance. The laboratory simulations reported here provide clear evidence for a solid-state mechanism leading to the formation of biologically relevant peptides under extraterrestrial conditions. This finding has important implications for the inventory of organic material inherited by young planetary bodies. Future studies should explore the limits of molecular complexity that can be achieved through such processes. In particular, irradiation temperature and dilution of glycine within water-rich ices may significantly influence both the yields and the types of products formed. It will also be important to assess the survivability of extraterrestrial peptides during their journey through space and delivery to planetary surfaces in order to fully evaluate their potential role in the origins of life on Earth.

\section{Methods}

The \textit{in situ} experimental work described in this article was carried out at the HUN-REN Institute for Nuclear Research (Atomki) in Debrecen, Hungary, on two ultra-high vacuum (UHV) end stations connected to separate ion sources. \textit{Ex situ} post-irradiation experiments were conducted using electrospray ionisation mass spectrometry (ESI-MS) in the Department of Molecular Biology and Genetics at Aarhus University in Denmark. 

\subsection*{\textit{In situ} ion irradiation experiments}
\subsubsection*{Atomki-Queen's University Ice Laboratory for Astrochemistry}\label{AQUILA}
The Atomki-Queen's University Ice Laboratory for Astrochemistry (AQUILA) at Atomki was used to expose three samples of glycine isomers (pure natural, partially deuterated, and fully deuterated glycine) to 10~keV protons at 20~K \cite{Aquilla}. The experimental setup is a cylindrical UHV compatible stainless steel chamber with an inner diameter of 300~mm and a height of 210~mm. The chamber is pumped down using three turbo molecular pumps (two Balzers TMU 260 pumps and a Varian TV301 pump) and one scroll pump (Edwards nXds 15i), which results in base pressures of $5\times10^{-9}$~mbar without cooling and better than $1\times10^{-9}$~mbar with cooling. In the centre of the chamber is a copper sample holder in which an infrared transparent zinc-selenide (\ce{ZnSe}) window of diameter 15~mm and thickness 3~mm is placed to act as a substrate. The sample holder can rotate about the vertical axis of the chamber using a 360° rotation stage and is in contact with the cold finger of a closed-cycle helium cryostat (Sumitomo CH-204SB-N cold head with a Sumitomo HC-4E1 compressor unit) that is used to cool the sample to a base temperature of 20~K \cite{Aquilla}. The sample temperature can then be controlled in a range of 20--300~K by using a proportional-integral-differential (PID) controller (Lake Shore model 336) connected to a 50~$\Omega$/100~W cartridge heater and three silicon diodes (Lake Shore DT-670B-CO). The chamber is equipped with a transmission FTIR Bruker Vertex V70v spectrometer and an external mercury-cadmium-telluride (MCT) detector cooled by liquid nitrogen to collect IR spectra of the ice sample in transmission mode. The initial step of all experiments was to cool the substrate to 20~K, flash heat it to 200~K to clean its surface of any potential contaminant such as residual water, and then cool it down again to 20~K. Next, a background FTIR measurement of the bare \ce{ZnSe} window was acquired at 20~K with a resolution of 1~cm$^{-1}$ and used as a reference spectrum for all the following measurements that used the same resolution. At this point, the glycine molecules were deposited on the substrate using a removable effusive evaporator (Createc OLED-40-10-WK-SHM) kept at 120 \textdegree C for approximately 20 minutes to allow the glycine powder located in the crucible of the effusive source to evaporate and uniformly re-condense on the cold \ce{ZnSe} window. During deposition, the sample holder was rotated to directly face the effusive evaporator. An FTIR spectrum of the deposited sample was acquired after deposition as shown in Fig.~\ref{Dep+irr}a. 

The AQUILA end station is connected to the Atomki electron cyclotron resonance (ECR) ion source, which is described in further detail in Ref.~\cite{2021EPJP..136..247B,Rcz2012MolecularAN}. The ECR ion source is capable of providing various low-energy ion beams of 5--20~keV singly charged ions, as well as multiply charged ions having energies of up to a few hundred keV. Ion beams are produced by generating the plasma state of the required material from its precursor gas via GHz-frequency microwave injection into the plasma chamber of the ion source. Different ions can be produced, such as \ce{P+}, \ce{Ca+}, \ce{Fe+}, \ce{Ni+}, \ce{Ag+}, \ce{Au+}, and \ce{C60} \cite{Aquilla}. The ions produced for the bombardment in this experiment were 10~keV protons. The AQUILA main chamber is separated from the ion beamline with a pneumatic gate valve and a differential pumping stage. The ion irradiation of the deposited glycine was performed at increasing fluences with an FTIR spectrum of the sample recorded after each step to monitor the physicochemical changes in the ice. When negligible changes in the IR spectra were observed, the irradiation was stopped at a total fluence of $6\times10^{15}$ protons, and a final FTIR spectrum was measured. After this, a temperature-programmed desorption (TPD) experiment was performed by heating the substrate at a ramp rate of 10~K/minute with IR spectra taken every 10~K. During TPD experiments, the AQUILA quadrupole mass spectrometer was used to measure molecules desorbing from the ice. During heating, the QMS measured m/z = 18, 19, and 20 to detect any desorption of \ce{H2O}, \ce{HDO} or \ce{D2O} from the surface. Once the \ce{ZnSe} window reached room temperature, the TPD was complete, the chamber was vented and the sample was removed and stored in a nitrogen atmosphere to avoid contamination from biological sources before further \textit{ex situ} analysis was performed using the ESI-MS setup described in the following.

\subsubsection*{Ice Chamber for Astrophysics-Astrochemistry}\label{ICA}
The 1~MeV proton bombardment irradiations were performed in the Ice Chamber for Astrophysics-Astrochemistry (ICA) \cite{2021RScI...92h4501H,Dunkpap}. The setup consists of a UHV compatible spherical decagon chamber with an inner diameter of 160 mm. A dry rough vacuum pump and a turbo molecular pump produce a base pressure of $10^{-8}$ mbar without cooling and $10^{-9}$ mbar with cooling. In the centre of the chamber there is a heat-shielded copper sample holder which allowed the mounting of three IR transparent \ce{ZnSe} windows. These windows act as the substrates on which the molecules, in this case the glycine isomers, can be deposited. Each window allows for different independent irradiations to take place and the sample holder itself can be moved via both rotations and in the vertical axis. This is done via a 360° rotational stage and a z-linear manipulator. The sample and substrates were cooled to 20~K using a closed-cycle helium cryostat (Leybold Coolpower 7/25 with a Leybold Coolpak 4000 compressor unit). The temperature was measured using two silicon diodes (Lake Shore DT-670B-CO) connected to a cryogenic temperature controller (Lake Shore model 335). The temperature of the sample can be regulated by heating with an internal 25 $\Omega$/100 W cartridge heater (HTR-25-100). This allows the temperature to vary between 20 and 300~K and can be linearly increased to allow TPD measurements to be taken \cite{2021RScI...92h4501H}. Molecules were deposited on the sample via a Knudsen cell evaporation source. This molecular doser is temperature controlled and therefore can be set to the required temperature for a molecular powder to evaporate. All experiments in this report that used this cell used a temperature of 110--125~°C. In the chamber there are two instruments that are used to measure the changes in the sample, an FTIR spectrometer and a QMS with a mass range of 1 to 300~amu. The IR spectrometer is a FTIR Bruker V70v with FIR extension with MCT and DLaTGS detectors in transmission mode. Experiments were carried out following the same procedure used at AQUILA, namely, cooling the substrates to 20~K, flash heating them to 200~K, cooling to 20~K again and acquiring a first IR reference spectrum with a resolution of 1~cm$^{-1}$ used as background for all the reminder spectra with the same resolution. 

The ICA end station operates in the Atomki Tandetron Laboratory, which is described in further detail here \cite{2021RScI...92h4501H,Rajta2018AcceleratorCO}. The facility hosts an ion accelerator that allows the production of protons and heavier ions in a range of energies. For protons, it can produce a beam with a range of energies of 0.2--4~MeV and beam currents of $>$200~$\mu$A \cite{2021RScI...92h4501H}. To ensure that the beam is stable and homogeneous on the surface of the sample, an x- and y-ion beam splitter is used to scan an area of 25~$\times$~25~mm. The current of the beam is continuously measured using Faraday cups along the beam path \cite{2021RScI...92h4501H}. For the experiments in this paper that used the ICA chamber connected to the Tandetron, the proton energies were set at 1~MeV. The ICA chamber is separated from the ion beamline, which operates at $10^{-6}$~mbar, by a series of gate valves and a differential pumping stage. This reduces the effect of contamination from the beamline into the ICA chamber. As in the case of AQUILA, the FTIR spectra were acquired after ice deposition and at every ion irradiation step. A TPD with a heat ramp of 10~K/minute was performed at the end of the irradiation experiment with IR spectra taken every 10~K. The residue obtained at 300~K was finally analysed \textit{in vacuo} using the FTIR spectrometer.

\subsection*{\textit{Ex situ} electrospray ionisation mass spectrometry}\label{ESI-MS_setup}
The end product after 10~keV proton irradiation of solid glycine isomers was collected from the ZnSe windows by washing them with 100 µl liquid chromatography - mass spectrometry (LC-MS) grade 0.1\% formic acid in water using a 200 µL micropipette with single use tips. The wash solution containing dissolved irradiation products was collected in Eppendorf tubes and immediately analysed by direct-infusion electrospray ionisation mass spectrometry (ESI-MS). An Orbitrap Eclipse mass spectrometer (Thermo Fisher Scientific) with a standard atmospheric pressure ionisation (API) source was used for the ESI-MS analysis. Direct infusion was performed using a syringe pump at a flow rate of 5 µL/min. The mass spectrometer was operated in positive-ion mode, with data acquisition over a mass range of 50 to 500 m/z. The resolution was set to 500.000, and the scan mode was Fourier Transform Mass Spectrometry (FTMS) + ESI full MS. For tandem mass spectrometry experiments (MS/MS), precursor ions were manually selected with a 2 Da isolation window, and fragmentation was achieved using higher-energy collisional dissociation (HCD) with collision energies set between 25 and 45. As can be seen in the figures, the masses shown are in four decimal places. The inability to calibrate exactly with the molecules detected meant that it was calibrated only to glycine. Therefore, the precision of the measured molecules may not be accurate to the theoretically calculated mass. However, because all samples are calibrated to the same point, they can be compared with each other as demonstrated with molecules that are expected to have the same m/z, such as glycine, being exactly the same.

\subsection*{Theoretical IR spectrum calculations}\label{Theory}

All theoretical calculations were performed with density functional theory (DFT) as implemented in the ORCA software package~\cite{ORCA,ORCA5}. An empirically weighted hybrid functional was used to model the electron exchange correlation interaction, consisting of a mixture of Becke's~\cite{becke1993density} three-parameter exchange functional with the Lee-Yang-Parr correlation functional~\cite{lee1988development} (B3LYP). We employed a def2-TZVP~\cite{weigend2005balanced} basis set and a def2/J~\cite{weigend2006accurate} auxiliary basis set. Dispersion effects were taken into account according to the Grimme D3 method~\cite{grimme2011effect,grimme2010consistent}. Before IR calculation, structural geometry optimization was performed at an appropriately high force convergence threshold of 0.005 eV\slash\r{A}, under the conductor-like polarizable continuum model (CPCM) to account for polar molecular crystal conditions. Molecular IR spectra were obtained through analytical Hessian calculation. IR spectra have been scaled according to the exchange correlation functional and the set of bases as suggested in Ref.~\cite{kashinski2017harmonic}.

\backmatter

\section{Data Availability}

Data for this project is hosted on Zenodo with the DOI:https://doi.org/10.5281/zenodo.17814722

Requests for materials should be addressed to Alfred Thomas Hopkinson* and Sergio Ioppolo**. 

\section{Acknowledgments}

Support from the Danish National Research Foundation through the Centre of Excellence 'InterCat' (grant agreement No. DNRF150) is acknowledged by A.T.H., A.M.W., J.P., A.T.M., L.H., and S.I. The research was supported by the Europlanet 2024 RI, which was funded by the European Union Horizon 2020 Research Innovation Programme under grant agreement No. 871149 and is acknowledged by A.T.H., N.J.M. The main components of the ICA setup were purchased using funds obtained from the Royal Society through grants UF130409, RGF/EA/180306, and URF/R/191018 and are acknowledged by S.I. Further developments of the installation were supported in part by the Eötvös Loránd Research Network through grants ELKH IF-2/2019 and ELKH IF-5/2020, and are acknowledged by R.R., P.H., G.L., S.B., Z.J., D.V.M., B.S., R.W.M., and N.J.M. This work has also received support from the European Union and the State of Hungary; cofinanced by the European Regional Development Fund through grant GINOP-2.3.3-15-2016-00005. Support has also been received from the Research, Development, and Innovation Fund of Hungary through grant nos. K128621 and ADVANCED-151196. These grants are acknowledged by R.R., P.H., G.L., S.B., Z.J., D.V.M., B.S., R.W.M., and N.J.M. This paper is also based on work from the COST Actions CA20129 “MultIChem” and CA22133 “PLANETS”, supported by COST (European Cooperation in Science and Technology) and are acknowledged by R.R., P.H., G.L., S.B., Z.J., D.V.M., B.S., R.W.M., and N.J.M. Zoltán Juhász is grateful for the support of the Hungarian Academy of Sciences through the János Bolyai Research Scholarship. Robert W. McCullough is the grateful recipient of an honorary visiting scholar post at Queen’s University Belfast.

\section{Author contributions statement}

A.T.H. conceived the experiments with input from S.I. and L.H.. A.T.H., A.M.W., and A.T.M. conducted the IR and QMS experiments, and A.T.H. analysed the results with input from S.I. and L.H.. R.R., P.H., G.L., S.B., Z.J., D.V.M., B.S., R.W.M., and N.J.M. enabled the irradiation experiments at HUN-REN Atomki. C.S. conducted the \textit{ex situ} ESI-MS experiments, and A.T.H. and C.S. analysed the results. J.P. conducted the theoretical calculations, and A.T.H. and J.P. analysed the results. A.T.H., C.S., J.P., and S.I. wrote the manuscript draft. All authors reviewed the manuscript. 

\section{Competing interests statement}

There are no known competing interests.

\bibliography{sample}% common bib file

@ARTICLE{1995Sci...270.1455S,
       author = {{Snow}, Theodore P. and {Witt}, Adolf N.},
        title = "{The Interstellar Carbon Budget and the Role of Carbon in Dust and Large Molecules}",
      journal = {Science},
         year = 1995,
        
       volume = {270},
       number = {5241},
        pages = {1455-1460},
          doi = {10.1126/science.270.5241.1455}
}

@ARTICLE{2014FaDi..168..267M,
       author = {{Mat{\'e}}, Bel{\'e}n and {Tanarro}, Isabel and {Moreno}, Miguel A. and {Jim{\'e}nez-Redondo}, Miguel and {Escribano}, Rafael and {Herrero}, V{\'\i}ctor J.},
        title = "{Stability of carbonaceous dust analogues and glycine under UV irradiation and electron bombardment}",
      journal = {Faraday Discuss.},
         year = 2014,
        
       volume = {168},
        pages = {267},
          doi = {10.1039/C3FD00132F}
}

@ARTICLE{2022FrASS...8..255P,
       author = {{Barone}, Vincenzo and {Puzzarini}, Cristina},
        title = "{Toward accurate formation routes of complex organic molecules in the interstellar medium: the paradigmatic cases of acrylonitrile and cyanomethanimine}",
      journal = {Front. Astron. Space Sci.},
         year = 2022,
        
       volume = {8},
          eid = {255},
        pages = {255},
          doi = {10.3389/fspas.2021.814384}
}

@article{doi.org/10.1002/cplu.202300492,
author = {Moreno, Abel and Bonduelle, Colin},
title = {New Insights on the Chemical Origin of Life: The Role of Aqueous Polymerization of {N}-carboxyanhydrides {(NCA)}},
journal = {ChemPlusChem},
volume = {89},
number = {7},
pages = {e202300492},
doi = {https://doi.org/10.1002/cplu.202300492},

eprint = {https://chemistry-europe.onlinelibrary.wiley.com/doi/pdf/10.1002/cplu.202300492},
year = {2024}
}

@article{mate,
author = {Maté, Belén and Tanarro, I. and Escribano, R. and Moreno Alba, Miguel and Herrero, V.},
year = {2015},

pages = {151},
title = {Stability of extraterrestrial glycine under energetic particle radiation estimated from 2 {keV} electron bombardment experiments},
volume = {806},
journal = {Astrophys. J.},
doi = {10.1088/0004-637X/806/2/151}
}

@article{howdep,
author = {Maté, Belén and Rodriguez, Yamilet and Gálvez, Oscar and Tanarro, I. and Escribano, Rafael},
year = {2011},

pages = {12268-12276},
title = {An infrared study of solid glycine in environments of astrophysical relevance},
volume = {13},
journal = {Phys. Chem. Chem. Phys.},
doi = {10.1039/c1cp20899c}
}

@article{protom,
   title={The Influence of Crystallinity Degree on the Glycine Decomposition Induced by 1 {MeV} Proton Bombardment in Space Analog Conditions},
   volume={13},
   ISSN={1557-8070},
  
   DOI={10.1089/ast.2012.0877},
   number={1},
   journal={Astrobiology},
   publisher={Mary Ann Liebert Inc},
   author={Pilling, Sergio and Mendes, Luiz A.V. and Bordalo, Vinicius and Guaman, Christian F.M. and Ponciano, Cássia R. and da Silveira, Enio F.},
   year={2013},
   month=jan, pages={79–91} }

@ARTICLE{2014MNRAS.441.3209P,
       author = {{Portugal}, Williamary and {Pilling}, Sergio and {Boduch}, Philippe and {Rothard}, Hermann and {Andrade}, Diana P.~P.},
        title = "{Radiolysis of amino acids by heavy and energetic cosmic ray analogues in simulated space environments: {\ensuremath{\alpha}}-glycine zwitterion form}",
      journal = {Mon. Not. R. Astron. Soc.},
         year = 2014,
        
       volume = {441},
       number = {4},
        pages = {3209-3225},
          doi = {10.1093/mnras/stu656},
archivePrefix = {arXiv},
       eprint = {1404.0894},
 primaryClass = {astro-ph.EP}
}

@article{Kaiser_2013,
doi = {10.1088/0004-637X/765/2/111},

year = {2013},

publisher = {The American Astronomical Society},
volume = {765},
number = {2},
pages = {111},
author = {R. I. Kaiser and A. M. Stockton and Y. S. Kim and E. C. Jensen and R. A. Mathies},
title = {On the formation of dipeptides in interstellar model ices},
journal = {Astrophys. J.}
}

@ARTICLE{2022NatAs...6..381K,
       author = {{Krasnokutski}, S.~A. and {Chuang}, K.-J. and {J{\"a}ger}, C. and {Ueberschaar}, N. and {Henning}, Th.},
        title = "{A pathway to peptides in space through the condensation of atomic carbon}",
      journal = {Nat. Astron.},
         year = 2022,
        
       volume = {6},
        pages = {381-386},
          doi = {10.1038/s41550-021-01577-9},
archivePrefix = {arXiv},
       eprint = {2202.12170},
 primaryClass = {physics.bio-ph}
}

@ARTICLE{2021RScI...92h4501H,
       author = {{Herczku}, P{\'e}ter and {Mifsud}, Duncan V. and {Ioppolo}, Sergio and {Juh{\'a}sz}, Zolt{\'a}n and {Ka{\r{A}}uchov{\'a}}, Zuzana and {Kov{\'a}cs}, S{\'a}ndor T.~S. and {Traspas Mui{\~n}a}, Alejandra and {Hailey}, Perry A. and {Rajta}, Istv{\'a}n and {Vajda}, Istv{\'a}n and {Mason}, Nigel J. and {McCullough}, Robert W. and {Parip{\'a}s}, B{\'e}la and {Sulik}, B{\'e}la},
        title = "{The Ice Chamber for Astrophysics-Astrochemistry (ICA): A new experimental facility for ion impact studies of astrophysical ice analogs}",
      journal = {Rev. Sci. Instrum.},
         year = 2021,
        
       volume = {92},
       number = {8},
          eid = {084501},
        pages = {084501},
          doi = {10.1063/5.0050930},
archivePrefix = {arXiv},
       eprint = {2109.11670},
 primaryClass = {astro-ph.IM}
}

@article{Rajta2018AcceleratorCO,
  title={Accelerator characterization of the new ion beam facility at {MTA Atomki in Debrecen, Hungary}},
  author={Istv{\'a}n Rajta and Istv{\'a}n Vajda and Gy. Gy{\"u}rky and L{\'a}szl{\'o} Csedreki and {\'A}rp{\'a}d Z. Kiss and S{\'a}ndor Biri and H. Oosterhout and Nicolae C. Podaru and Dirk J. W. Mous},
    doi = {https://doi.org/10.1016/j.nima.2017.10.073},
  journal={Nucl. Instrum. Methods Phys. Res., Sect. A},
  year={2018},
  volume={880},
  pages={125-130},
 
}

@ARTICLE{2021EPJP..136..247B,
       author = {{Biri}, S. and {Vajda}, I.~K. and {Hajdu}, P. and {R{\'a}cz}, R. and {Cs{\'\i}k}, A. and {Korm{\'a}ny}, Z. and {Perduk}, Z. and {Kocsis}, F. and {Rajta}, I.},
        title = "{The Atomki Accelerator Centre}",

      journal = {Eur. Phys. J. Plus},
         year = 2021,
        
       volume = {136},
       number = {2},
          eid = {247},
        pages = {247},
          doi = {10.1140/epjp/s13360-021-01219-z}
}

@article{Rcz2012MolecularAN,
  title={Molecular and negative ion production by a standard electron cyclotron resonance ion source.},
  author={Rich{\'a}rd R{\'a}cz and S{\'a}ndor Biri and Zolt{\'a}n Juh{\'a}sz and B{\'e}la Sulik and J{\'o}zsef P{\'a}link{\'a}s},
  journal={Rev. Sci. Instrum.},
  year={2012},
  volume={83 2},
  pages={
          02A313
        },

}

@article{Aquilla,
    author = {Rácz, R. and Kovács, S. T. S. and Lakatos, G. and Rahul, K. K. and Mifsud, D. V. and Herczku, P. and Sulik, B. and Juhász, Z. and Perduk, Z. and Ioppolo, S. and Mason, N. J. and Field, T. A. and Biri, S. and McCullough, R. W.},
    title = {{AQUILA}: A laboratory facility for the irradiation of astrochemical ice analogs by {keV} ions},
doi = {10.1063/1.3662960},
    journal = {Rev. Sci. Instrum.},
    volume = {95},
    number = {9},
    pages = {095105},
    year = {2024},
    issn = {0034-6748},

   
    eprint = {https://pubs.aip.org/aip/rsi/article-pdf/doi/10.1063/5.0207967/20146934/095105\_1\_5.0207967.pdf},
}

@ARTICLE{2021NatAs...5..197I,
       author = {{Ioppolo}, S. and {Fedoseev}, G. and {Chuang}, K. -J. and {Cuppen}, H.~M. and {Clements}, A.~R. and {Jin}, M. and {Garrod}, R.~T. and {Qasim}, D. and {Kofman}, V. and {van Dishoeck}, E.~F. and {Linnartz}, H.},
        title = "{A non-energetic mechanism for glycine formation in the interstellar medium}",
      journal = {Nat. Astron.},
         year = 2021,
        
       volume = {5},
        pages = {197-205},
          doi = {10.1038/s41550-020-01249-0},
archivePrefix = {arXiv},
       eprint = {2011.06145},
primaryClass = {astro-ph.IM}
}

@article{doi:10.1021/cr2004844,
author = {Ruiz-Mirazo, Kepa and Briones, Carlos and de la Escosura, Andrés},
title = {Prebiotic Systems Chemistry: New Perspectives for the Origins of Life},
journal = {Chem. Rev.},
volume = {114},
number = {1},
pages = {285-366},
year = {2014},
doi = {10.1021/cr2004844},
    note ={PMID: 24171674},


eprint = {        https://doi.org/10.1021/cr2004844}
    
}

@ARTICLE{2019A&A...630A..32H,
       author = {{Hadraoui}, K. and {Cottin}, H. and {Ivanovski}, S.~L. and {Zapf}, P. and {Altwegg}, K. and {Benilan}, Y. and {Biver}, N. and {Della Corte}, V. and {Fray}, N. and {Lasue}, J. and {Merouane}, S. and {Rotundi}, A. and {Zakharov}, V.},
        title = "{Distributed glycine in comet 67P/Churyumov-Gerasimenko}",
      journal = {Astron. Astrophys.},
         year = 2019,
        
       volume = {630},
          eid = {A32},
        pages = {A32},
          doi = {10.1051/0004-6361/201935018}
}

@ARTICLE{2005RSPTA.363.2729S,
       author = {{Sephton}, Mark A.},
        title = "{Organic matter in carbonaceous meteorites: past, present and future research}",
      journal = {Philos. Trans. R. Soc. London, Ser. A},
         year = 2005,
        
       volume = {363},
       number = {1837},
        pages = {2729-2742},
          doi = {10.1098/rsta.2005.1670}
}

@ARTICLE{2015Sci...347a0628C,
       author = {{Capaccioni}, F. and {Coradini}, A. and {Filacchione}, G. and {Erard}, S. and {Arnold}, G. and {Drossart}, P. and {De Sanctis}, M.~C. and {Bockelee-Morvan}, D. and {Capria}, M.~T. and {Tosi}, F. and {Leyrat}, C. and {Schmitt}, B. and {Quirico}, E. and {Cerroni}, P. and {Mennella}, V. and {Raponi}, A. and {Ciarniello}, M. and {McCord}, T. and {Moroz}, L. and {Palomba}, E. and {Ammannito}, E. and {Barucci}, M.~A. and {Bellucci}, G. and {Benkhoff}, J. and {Bibring}, J.~P. and {Blanco}, A. and {Blecka}, M. and {Carlson}, R. and {Carsenty}, U. and {Colangeli}, L. and {Combes}, M. and {Combi}, M. and {Crovisier}, J. and {Encrenaz}, T. and {Federico}, C. and {Fink}, U. and {Fonti}, S. and {Ip}, W.~H. and {Irwin}, P. and {Jaumann}, R. and {Kuehrt}, E. and {Langevin}, Y. and {Magni}, G. and {Mottola}, S. and {Orofino}, V. and {Palumbo}, P. and {Piccioni}, G. and {Schade}, U. and {Taylor}, F. and {Tiphene}, D. and {Tozzi}, G.~P. and {Beck}, P. and {Biver}, N. and {Bonal}, L. and {Combe}, J. -Ph. and {Despan}, D. and {Flamini}, E. and {Fornasier}, S. and {Frigeri}, A. and {Grassi}, D. and {Gudipati}, M. and {Longobardo}, A. and {Markus}, K. and {Merlin}, F. and {Orosei}, R. and {Rinaldi}, G. and {Stephan}, K. and {Cartacci}, M. and {Cicchetti}, A. and {Giuppi}, S. and {Hello}, Y. and {Henry}, F. and {Jacquinod}, S. and {Noschese}, R. and {Peter}, G. and {Politi}, R. and {Reess}, J.~M. and {Semery}, A.},
        title = "{The organic-rich surface of comet 67P/Churyumov-Gerasimenko as seen by VIRTIS/Rosetta}",
      journal = {Science},
         year = 2015,
        
       volume = {347},
       number = {6220},
          eid = {aaa0628},
        pages = {aaa0628},
          doi = {10.1126/science.aaa0628}
}

@article{article,
author = {Esmaili, Sasan and Bass, Andrew and Cloutier, Pierre and Sanche, Léon and Huels, Michael},
year = {2018},

pages = {164702},
title = {Glycine formation in {CO2 :CH4 :NH3} ices induced by 0-70 {eV} electrons},
volume = {148},
journal = {J. Chem. Phys.},
doi = {10.1063/1.5021596}
}

@ARTICLE{2018MNRAS.479..130U,
       author = {{Urso}, R.~G. and {Palumbo}, M.~E. and {Baratta}, G.~A. and {Scir{\`e}}, C. and {Strazzulla}, G.},
        title = "{Solid deuterated water in space: detection constraints from laboratory experiments}",
      journal = {Mon. Not. R. Astron. Soc.},
         year = 2018,
        
       volume = {479},
       number = {1},
        pages = {130-140},
          doi = {10.1093/mnras/sty1428}
}

@ARTICLE{Gorlero2009153,
	author = {Gorlero, Maçha and Wieczorek, Rafal and Adamala, Katarzyna and Giorgi, Alessandra and Schininà, Maria Eugenia and Stano, Pasquale and Luisi, Pier Luigi},
	title = {{Ser-His} catalyses the formation of peptides and {PNAs}},
	year = {2009},
	journal = {FEBS Lett.},
	volume = {583},
	number = {1},
	pages = {153 – 156},
	doi = {10.1016/j.febslet.2008.11.052},

	type = {Article},
	publication_stage = {Final},
	source = {Scopus},
	note = {Cited by: 77; All Open Access, Bronze Open Access}
}

@article{https://doi.org/10.1002/anie.201000212,
author = {Childers, W. Seth and Mehta, Anil K. and Ni, Rong and Taylor, Jeannette V. and Lynn, David G.},
title = {Peptides Organized as Bilayer Membranes},
journal = {Angew. Chem. Int. Ed.},
volume = {49},
number = {24},
pages = {4104-4107},
doi = {https://doi.org/10.1002/anie.201000212},
year = {2010}
}

@article{doi:10.1080/07391102.2022.2088619,
author = {Prakash Kulkarni, Ravi Salgia and Vladimir N. Uversky},
title = {Intrinsic disorder, extraterrestrial peptides, and prebiotic life on the earth},
journal = {J. Biomol. Struct. Dyn.},
volume = {41},
number = {12},
pages = {5481--5485},
year = {2023},
publisher = {Taylor \& Francis},
doi = {10.1080/07391102.2022.2088619},

    note ={PMID: 35723592},

}

@article{doi:10.1098/rsif.2021.0641,
author = {Fried, Stephen D.  and Fujishima, Kosuke  and Makarov, Mikhail  and Cherepashuk, Ivan  and Hlouchova, Klara },
title = {Peptides before and during the nucleotide world: an origins story emphasizing cooperation between proteins and nucleic acids},
journal = {J. R. Soc. Interface},
volume = {19},
number = {187},
pages = {20210641},
year = {2022},
doi = {10.1098/rsif.2021.0641}
}

@article{doi:10.1021/cr400153k,
author = {Bennett, Chris J. and Pirim, Claire and Orlando, Thomas M.},
title = {Space-Weathering of Solar System Bodies: A Laboratory Perspective},
journal = {Chem. Rev.},
volume = {113},
number = {12},
pages = {9086-9150},
year = {2013},
doi = {10.1021/cr400153k},
    note ={PMID: 24274796},


}

@article{10.1093/mnras/staa3939,
    author = {da Costa, C A P and Souza-Corrêa, J A and da Silveira, E F},
    title = "{Infrared analysis of Glycine dissociation by MeV ions and keV electrons}",
    journal = {Mon. Not. R. Astron. Soc.},
    volume = {502},
    number = {2},
    pages = {2105-2119},
    year = {2021},
    issn = {0035-8711},
    doi = {10.1093/mnras/staa3939},
}

@article{Uvamino,
author = {Munoz-Caro, Guillermo and Meierhenrich, Uwe and Schutte, W. and Barbier, B and Segovia, A. and Rosenbauer, F. and Thiemann, Wolfram and H.-P, Wolfram and Brack, Andre and Greenberg},
doi = {https://doi.org/10.1038/416403a},
year = {2002},

title = {Amino acids from ultraviolet irradiation of interstellar ice analogues},
volume = {416},
pages = {403-406},
journal = {Nature},
}

@article{anotherUVamino,
author = {Bernstein, Max and Dworkin, Jason and Sandford, Scott and Cooper, George and Allamandola, Louis},
year = {2002},

pages = {401-403},
title = {Racemic amino acids from the ultraviolet photolysis of interstellar ice analogues},
volume = {416},
journal = {Nature},
doi = {10.1038/416401a}
}

@article{Gerakines2012InSM,
  title={In situ measurements of the radiation stability of amino acids at 15–140 {K}},
  author={Perry A. Gerakines and Reggie L. Hudson and Marla H. Moore and Jan-Luca Bell},
  journal={Icarus},
  year={2012},
  volume={220},
  pages={647-659},
 doi = {https://doi.org/10.1016/j.icarus.2012.06.001},
}

@article{JHEETA2012208,
title = {The irradiation of 1:1 mixture of ammonia:carbon dioxide ice at {30 K} using {1 keV} electrons},
journal = {Chem. Phys. Lett.},
volume = {543},
pages = {208-212},
year = {2012},
issn = {0009-2614},
doi = {https://doi.org/10.1016/j.cplett.2012.06.051},
author = {S. Jheeta and S. Ptasinska and B. Sivaraman and N.J. Mason}
}

@article{doi:10.1021/bi800329p,
author = {Morar, Mariya and Hoskins, Aaron A. and Stubbe, JoAnne and Ealick, Steven E.},
title = {Formylglycinamide Ribonucleotide Amidotransferase from Thermotoga maritima: Structural Insights into Complex Formation},
journal = {Biochemistry},
volume = {47},
number = {30},
pages = {7816-7830},
year = {2008},
doi = {10.1021/bi800329p},

}

@article{PEDLEY2017141,
title = {A New View into the Regulation of Purine Metabolism: The Purinosome},
journal = {Trends Biochem. Sci.},
volume = {42},
number = {2},
pages = {141-154},
year = {2017},
issn = {0968-0004},
doi = {https://doi.org/10.1016/j.tibs.2016.09.009},
author = {Anthony M. Pedley and Stephen J. Benkovic}
}

@article{OD,
    title={Analysis of deuterated water contents using {FTIR} bending motion},
    author={Park, Kyueun and Kim, Youngjin and Lee, Kyung Jin},
    volume={322},
    doi={10.1007/s10967-019-06734-z},
    number={2},
    journal={J. Radioanal. Nucl. Chem.},
    publisher={Springer Science and Business Media LLC},
    year={2019},
    month={Sep},
    pages={487-493},
    language={en},
}

@article{PALUMBO201064,
title = {H bonds in astrophysical ices},
journal = {J. Mol. Struct.},
volume = {972},
number = {1},
pages = {64-67},
year = {2010},
issn = {0022-2860},
doi = {https://doi.org/10.1016/j.molstruc.2009.12.017},
author = {M.E. Palumbo and G.A. Baratta and G. Leto and G. Strazzulla}
}

@article{ORCA,
author = {Neese,F.},
title = {The {ORCA} program system},
journal = {WIRES Comput. Molec. Sci.},
volume = {2},
number = {1},
pages = {73-78},
DOI = {10.1002/wcms.81},
year = {2012},
type = {journal Article}
}

@article{ORCA5,
author = {Neese,F.},
title = {Software update: the {ORCA} program system, version 5.0},
journal = {WIRES Comput. Molec. Sci.},
volume = {12},
number = {1},
pages = {e1606},
DOI = {10.1002/wcms.1606},
year = {2022},
type = {journal Article}
}

@article{grimme2010consistent,
  title={A consistent and accurate ab initio parametrization of density functional dispersion correction {(DFT-D)} for the 94 elements {H-Pu}},
  author={Grimme, Stefan and Antony, Jens and Ehrlich, Stephan and Krieg, Helge},
  journal={J. Chem. Phys.},
  volume={132},
  number={15},
  year={2010},
DOI = {10.1063/1.3382344},

  publisher={AIP Publishing}
}

@article{grimme2011effect,
  title={Effect of the damping function in dispersion corrected density functional theory},
  author={Grimme, Stefan and Ehrlich, Stephan and Goerigk, Lars},
  journal={J. Comput. Chem.},
  volume={32},
  number={7},
  pages={1456--1465},
  year={2011},
  publisher={Wiley Online Library},
doi = {https://doi.org/10.1002/jcc.21759},

}

@article{becke1993density,
  title={Density-functional thermochemistry. III. The role of exact exchange},
  author={Becke, Axel D},
  journal={J. Chem. Phys.},
  volume={98},
  number={7},
  pages={5648--5652},
  year={1993},
  publisher={American Institute of Physics},
doi = {10.1063/1.464913},

}

@article{lee1988development,
  title={Development of the {Colle-Salvetti} correlation-energy formula into a functional of the electron density},
  author={Lee, Chengteh and Yang, Weitao and Parr, Robert G},
  journal={Phys. Rev. B: Condens. Matter},
  volume={37},
  number={2},
  pages={785},
  year={1988},
  publisher={APS},
doi = {10.1103/PhysRevB.37.785},
}

@article{weigend2005balanced,
  title={Balanced basis sets of split valence, triple zeta valence and quadruple zeta valence quality for {H to Rn} : Design and assessment of accuracy},
  author={Weigend, Florian and Ahlrichs, Reinhart},
  journal={Phys. Chem. Chem. Phys.},
  volume={7},
  number={18},
  pages={3297--3305},
  year={2005},
  publisher={Royal Society of Chemistry},
doi  ="10.1039/B508541A",
}

@article{weigend2006accurate,
  title={Accurate Coulomb-fitting basis sets for H to Rn},
  author={Weigend, Florian},
  journal={Phys. Chem. Chem. Phys.},
  volume={8},
  number={9},
  pages={1057--1065},
  year={2006},
  publisher={Royal Society of Chemistry},
doi  ="10.1039/B515623H",
}

@article{kashinski2017harmonic,
  title={Harmonic vibrational frequencies: approximate global scaling factors for {TPSS, M06, and M11} functional families using several common basis sets},
  author={Kashinski, DO and Chase, GM and Nelson, RG and Di Nallo, OE and Scales, AN and VanderLey, DL and Byrd, EFC},
  journal={J. Phys. Chem. A},
  volume={121},
  number={11},
  pages={2265--2273},
  year={2017},
  publisher={ACS Publications},
doi = {10.1021/acs.jpca.6b12147},
}

@article{wild2,
author = {Elsila, Jamie E. and Glavin, Daniel P. and Dworkin, Jason P.},
title = {Cometary glycine detected in samples returned by {Stardust}},
journal = {	Meteorit. Planet. Sci.},
volume = {44},
number = {9},
pages = {1323-1330},
doi = {https://doi.org/10.1111/j.1945-5100.2009.tb01224.x},
year = {2009}
}

@article{morewild,
author = {Scott A. Sandford  and Jérôme Aléon  and Conel M. O'D. Alexander  and Tohru Araki  and Sasa Bajt  and Giuseppe A. Baratta  and Janet Borg  and John P. Bradley  and Donald E. Brownlee  and John R. Brucato  and Mark J. Burchell  and Henner Busemann  and Anna Butterworth  and Simon J. Clemett  and George Cody  and Luigi Colangeli  and George Cooper  and Louis D'Hendecourt  and Zahia Djouadi  and Jason P. Dworkin  and Gianluca Ferrini  and Holger Fleckenstein  and George J. Flynn  and Ian A. Franchi  and Marc Fries  and Mary K. Gilles  and Daniel P. Glavin  and Matthieu Gounelle  and Faustine Grossemy  and Chris Jacobsen  and Lindsay P. Keller  and A. L. David Kilcoyne  and Jan Leitner  and Graciela Matrajt  and Anders Meibom  and Vito Mennella  and Smail Mostefaoui  and Larry R. Nittler  and Maria E. Palumbo  and Dimitri A. Papanastassiou  and François Robert  and Alessandra Rotundi  and Christopher J. Snead  and Maegan K. Spencer  and Frank J. Stadermann  and Andrew Steele  and Thomas Stephan  and Peter Tsou  and Tolek Tyliszczak  and Andrew J. Westphal  and Sue Wirick  and Brigitte Wopenka  and Hikaru Yabuta  and Richard N. Zare  and Michael E. Zolensky },
title = {Organics Captured from Comet {81P/Wild} 2 by the {Stardust} Spacecraft},
journal = {Science},
volume = {314},
number = {5806},
pages = {1720-1724},
year = {2006},
doi = {10.1126/science.1135841}}

@article{SZNAJDER20234923,
title = {Solar wind {H+} fluxes at 1 {AU} for solar cycles 23 and 24},
journal = {Adv. Space Res.},
volume = {71},
number = {11},
pages = {4923-4957},
year = {2023},
issn = {0273-1177},
doi = {https://doi.org/10.1016/j.asr.2023.01.054},
author = {Maciej Sznajder}
}

@article{shock,
author = {Martins, Zita and Price, Mark and Goldman, Nir and Sephton, Mark and Burchell, Mark},
year = {2013},

pages = {1045-1049},
title = {Shock synthesis of amino acids from impacting cometary and icy planet surface analogues},
volume = {6},
journal = {Nat. Geosci.},
doi = {10.1038/ngeo1930}
}

@article{singh2020shock,
  title={Shock processing of amino acids leading to complex structures—implications to the origin of life},
  author={Singh, Surendra V and Vishakantaiah, Jayaram and Meka, Jaya K and Sivaprahasam, Vijayan and Chandrasekaran, Vijayanand and Thombre, Rebecca and Thiruvenkatam, Vijay and Mallya, Ambresh and Rajasekhar, Balabhadrapatruni N and Muruganantham, Mariyappan and others},
  journal={Molecules},
  volume={25},
  number={23},
  pages={5634},
  year={2020},
  publisher={MDPI},
DOI = {10.3390/molecules25235634}

}

@article{Dunkpap,
	author = {Mifsud, Duncan V. and {Juhász, Zoltán} and {Herczku, Péter} and {Kovács, Sándor T. S.} and {Ioppolo, Sergio} and {Kaňuchová, Zuzana} and {Czentye, Máté} and {Hailey, Perry A.} and {Muiña, Alejandra Traspas} and {Mason, Nigel J.} and {McCullough, Robert W.} and {Paripás, Béla} and {Sulik, Béla}},
	title = {Electron irradiation and thermal chemistry studies of interstellar and planetary ice analogues at the {ICA} astrochemistry facility},
	DOI= "10.1140/epjd/s10053-021-00192-7",
	journal = {Eur. Phys. J. D},
	year = 2021,
	volume = 75,
	number = 6,
	pages = "182",
}

@article{newwater,
	author = {{Escribano, B.} and {del Burgo Olivares, C.} and {Carrascosa, H.} and {Cazaux, S.} and {Satorre, M. Á.} and {Muñoz Caro, G. M.}},
	title = {Interstellar water ice analogue properties as a function of temperature: Updated density, porosity, and infrared band strength},
	DOI= "10.1051/0004-6361/202555090",
	journal = {Astron. Astrophys.},
	year = 2025,
	volume = 699,
	pages = "A79",
}


\end{document}